\documentclass[conference]{IEEEtran}
\IEEEoverridecommandlockouts
\usepackage{cite}
\usepackage{amsmath,amssymb,amsfonts}
\usepackage{graphicx}
\usepackage{textcomp}
\usepackage{xcolor}
\def\BibTeX{{\rm B\kern-.05em{\sc i\kern-.025em b}\kern-.08em
\kern-.1667em\lower.7ex\hbox{E}\kern-.125emX}}

\usepackage{subfig}
\usepackage{graphicx}
\usepackage{graphics}
\usepackage{epstopdf}
\usepackage{amsmath}
\usepackage{multirow}
\usepackage{bm}
\usepackage{enumerate}
\usepackage{mathdots}
\usepackage{todonotes}
\usepackage{algorithmicx}
\usepackage{algpseudocode}
\usepackage{algorithm}
\usepackage{verbatim}
\usepackage{balance}
\usepackage{cite}
\usepackage{url}
\usepackage{acronym}
\usepackage{booktabs}
\usepackage{tabularx}

\newacro{ML}{Maximum Likelihood}

\usepackage{color}
\definecolor{purple(x11)}{rgb}{0.63, 0.36, 0.94}
\definecolor{cadmiumgreen}{rgb}{0.0, 0.42, 0.24}

\newcommand{\Real}{\mathop{\mathrm{Re}}}

\newcommand{\diag}{\mathop{\mathrm{diag}}}

\newcommand{\SNR}{\mathop{\mathrm{SNR}}}

\usepackage{pifont}

\newcommand{\Nt}{N_{\mathrm{t}}}

\newcommand{\be}{\begin{eqnarray}}
\newcommand{\ee}{\end{eqnarray}}

\def\bee{{\mathbf{e}}}
\def\bff{{\mathbf{f}}}

\def\bh{{\mathbf{h}}}

\def\bx{{\mathbf{x}}}

\def\b0{{\mathbf{0}}}

\def\bC{{\mathbf{C}}}

\def\bH{{\mathbf{H}}}

\def\sfj{{\mathsf{j}}}

\def\sf0{{\mathsf{0}}}

\def\bsf0{{\bm{\mathsf{0}}}}

\begin{document}
\title{Spectral Efficiency vs Complexity in Downlink Algorithms for Reconfigurable Intelligent Surfaces}
\author{Pooja Nuti $^{\dag}$ and Brian L. Evans $^{\dag}$\\
$^\dag$ The University of Texas at Austin, Email: $\{$pnuti@utexas.edu,bevans@ece.utexas.edu$\}$}
\maketitle

\begin{abstract}
Reconfigurable Intelligent Surfaces (RIS) are an emerging technology that can be used to reconfigure the propagation environment to improve cellular communication link rates. RIS, which are thin metasurfaces composed of discrete elements, passively manipulate incident electromagnetic waves through controlled reflective phase tuning.  In this paper, we investigate co-design of the multiantenna basestation beamforming vector and multielement RIS phase shifts. The downlink narrowband transmission uses sub-6 GHz frequency bands, and the user equipment has a single antenna. Subject to the non-convex constraints due to the RIS phase shifts, we maximize the spectral efficiency or equivalent channel power as a proxy.  Our contributions in improving RIS-aided links include  (1) design of gradient ascent codesign algorithms, and (2) comparison of seven codesign algorithms in spectral efficiency vs. computational complexity. In simulation, the best spectral efficiency vs. computational complexity tradeoffs are shown by two of our proposed gradient ascent algorithms.

\end{abstract}

\section{Introduction}
The global acceleration of 5G promises rapid increases in 5G connections and devices expected in the coming years. During this time, research has already began in 6G technologies which can address limitations of current solutions \cite{green_comm,what_will_5g_be}. In modern wireless networks the channel is frequently regarded as highly probabilistic and uncontrollable. Reconfigurable Intelligent Surfaces (RISs) or large intelligent surfaces (LISs) are an emerging technology expected for 6G which can reconfigure the propagation environment\cite{intelligent_walls_smart_environment_paper,metasurface_magazine,overview_RIS_journal,overview_RIS_magazine,overview_paper}. The primary enabling technology considered behind RISs are metasurfaces.

Metasurfaces are thin surfaces made of discrete elements which behave as passive reflectors of impinging signals; furthermore, the internal metasurface structure distinguishes them as materials in contrast to antenna arrays \cite{intelligent_walls_smart_environment_paper, metasurface_magazine, overview_paper}. An appealing aspect of RIS is their ability to manipulate the incident electromagnetic wave through controlled reflective phase tuning.


RIS research is gaining momentum because of its potential to manipulate the propagation environment favorably; however preceding technologies capable of improving the channel between transmitter and receiver include wireless repeaters, relays and reflect array antennas \cite{RIS_myths}. The distinctive aspects of RIS include that RISs are programmable, and do not require power amplifiers and complex processing, encoding and decoding algorithms \cite{RIS_vs_Relays}. Recently there has been various efforts to optimize various performance metrics in RIS-assisted communication systems \cite{survey_optimization_lis}. In this paper, we focus on maximizing the spectral efficiency for a single-user within a sub-6 GHz RIS-assisted communication system. 

Most of prior work on RIS focuses on narrowband MISO communication systems \cite{ee_opt_ris, indoor_signal_focusing_RIS, discrete_phase_SU_RIS, finite_element_iteration_paper, SDR_paper} 
Recently, several works have been made which focus on narrowband MISO channels, and attempt to configure both the RIS phase elements and beamformer through finding tractable suboptimal solutions optimizing different performance metrics, such as the energy efficiency \cite{ee_opt_ris}, and spectral efficiency, both in the single \cite{discrete_phase_SU_RIS, finite_element_iteration_paper, SDR_paper}  and multi user setting \cite{discrete_phase_SU_RIS, mu_rate_optimization_tHz, mu_sum_rate_miso_opt_RIS, multiuser_RIS_problem}.

In \cite{finite_element_iteration_paper}, the joint problem of beamformer and RIS elements optimization was formulated as a spectral efficiency maximization problem, for channels having both LOS and NLOS components. Owing to the difficulty of optimization under non-convex unit-magnitude constraints for the RIS elements, two solutions are proposed therein offering similar performance, the first one exploiting fixed point iteration and a more computationally complex Riemaniann manifold optimization. 
Another solution was proposed in \cite{SDR_paper} to design both the beamformer and RIS phase elements in a single user narrowband MISO setting by optimizing the received signal power subject to the non-convex hardware constraints from the RIS elements. In \cite{SDR_paper}, both a centralized and distributed algorithm are proposed which offer similar performance, the former using  semidefinite relaxation (SDR) requiring prohibitive signaling overhead, and the latter being a reduced complexity version of the former. The distributed method, however, requires the RIS to be able to have receive signal processing capabilities (i.e. features of channel estimation) and feedback estimated RIS phases to the AP, which updates the beamforming vector. 

The main challenge in optimizing the spectral efficiency in a RIS-assisted communication system comprises the non-convex unit-modular constraints on the RIS elements. To obtain tractable solutions to this non-convex problem, prior work employed alternate maximization and convex relaxation, among other techniques leading to sub optimal solutions \cite{discrete_phase_SU_RIS, finite_element_iteration_paper, SDR_paper}. As a consequence, optimizing throughput in RIS-assisted communication systems remains an open research problem.\\  
\indent In this paper we investigate a narrowband, downlink, single user equipped with a single antenna scenario in which a RIS is located in close proximity to a basestation. We propose several solutions to the problem of joint RIS phase elements and beamformer optimization, under a system model for sub-$6$ GHz communications
Prior work primarily optimizes the received signal power \cite{discrete_phase_SU_RIS, finite_element_iteration_paper, SDR_paper}, we distinctly consider direct optimization of the spectral efficiency function, which is a function of the SNR itself. In addition, we employ a projected gradient ascent method with initialization to enhance convergence of the proposed algorithms. The proposed approaches consist of the power method and projected gradient ascent methods both when optimizing over the received signal power and the spectral efficiency. In the numerical results section, we compare the proposed algorithms and those in \cite{finite_element_iteration_paper, SDR_paper} and show that the proposed approaches greatly outperform prior work solutions, both in terms of achievable spectral efficiency and computational complexity.

\section{System Model}
\subsection{Problem Formulation}
We consider a single user MISO communication system in which a transmitter equipped with $N_\text{t}$ antennas sends a data stream to a single-antenna receiver using both a direct LOS link and a RIS-aided NLOS link, the RIS being equipped with $N_\text{RIS}$ phase-shifters. The channel corresponding to the direct link between the transmitter and receiver is assumed to follow a Rayleigh channel model \cite{discrete_phase_SU_RIS,ee_opt_ris, indoor_signal_focusing_RIS,finite_element_iteration_paper, SDR_paper} and is modeled as $\bh_\text{d} \in \mathbb{C}^{1 \times N_\text{t}}$. The channel between the transmitter and the RIS surface is modeled as a Rayleigh matrix $\bH_1 \in \mathbb{C}^{N_\text{RIS} \times N_\text{t}}$, and the channel between the RIS and the receiver is modeled as $\bh_2 \in \mathbb{C}^{1 \times N_\text{RIS}}$ using a Rayleigh channel model as well. The RIS behavior is modeled as $\bm \Phi \in \mathbb{C}^{N_\text{RIS} \times N_\text{RIS}}$, $\bm \Phi = \diag\{ e^{\sfj \phi_1}, \ldots, e^{\sfj \phi_{N_\text{RIS}}} \}$. We define an equivalent channel comprises of the LOS and NLOS paths between the BS and UE as follows
\begin{equation}
\mathbf{h}_\text{eq} = \mathbf{h}_\text{d} + \mathbf{h}_2 \mathbf{\Phi} \mathbf{H}_1
\label{equation:h_eq}
\end{equation}
The received signal can then be expressed as
\begin{equation}
	y = \mathbf{h}_\text{eq} \bff s + w,
	\label{equation:rx_signal_general}
\end{equation}
in which $\bff \in \mathbb{C}^{N_\text{t} \times 1}$ is the transmit beamforming vector normalized so that $\|\bff\|_2^2 = 1$, $s \in \mathbb{C}$ is the transmit symbol satisfying $\mathbb{E}\{ |s|^2 \} = 1$, and $w \in \mathbb{C}$ is the receive circularly-symmetric additive white Gaussian noise sample $w \sim {\cal CN}(0,\sigma_w^2)$. The SNR is defined as $\text{SNR} = \frac{1}{\sigma_w^2}$.
In this paper, we focus on the joint problem of finding the beamformer $\bff$ and RIS matrix $\bm \Phi$ that optimize the resulting spectral efficiency. Let $\bh_\text{eq} \in \mathbb{C}^{1 \times \Nt}$ be the equivalent channel including both the LOS and NLOS paths as $\mathbf{h}_\text{eq} =\left(\bh_\text{d} + \bh_2 \bm \Phi \bH_1 \right)$. The problem of finding $\bff$ and $\bm \Phi$ can be formalized as
\begin{equation}
\begin{split}
	\left\{\bff^\star, \bm \Phi^\star\right\} =\,\, &\underset{\bff, \bm \Phi}{\arg\,\max}\, {\cal R}(\bff,\bm \Phi) \\
	&\text{subject to } \left| \bee_i^T \bm \Phi \bee_i \right|^2 = 1, \quad i = 1 \ldots N_\text{RIS}.
\end{split}
\label{equation:opt_problem_spectral_efficiency}
\end{equation}
 From \eqref{equation:rx_signal_general}, since the transmitted symbol is Gaussian, and the receive additive noise is also Gaussian, it follows that the spectral efficiency ${\cal R}(\bff,\bm \Phi)$ in \eqref{equation:opt_problem_spectral_efficiency} is given by
\begin{equation}
	{\cal R}(\bff,\bm \Phi) = \log_2 \left( 1 + \SNR \|\left(\bh_\text{d} + \bh_2 \bm \Phi \bH_1 \right) \bff\|_2^2 \right).
	\label{equation:spectral_efficiency_general}
\end{equation}
The spectral efficiency in \eqref{equation:spectral_efficiency_general} implicitly assumes a MISO communication link. To build intuition into the nature and solution to \eqref{equation:opt_problem_spectral_efficiency}, in the next section we will first briefly review the SISO optimization, and then we will investigate the MISO optimization.

\section{Application Cases}
\subsection{Single-User SISO with multiple RIS elements}
In \cite{for_siso_setting}, the simplest scenario of configuring the RIS elements to maximize spectral efficiency is considered for the single-user (SU) case. In the event that $N_\text{RIS} >1$, the resulting optimized RIS elements are a straightforward extension from the case of $N_\text{RIS} = 1$. The spectral efficiency for the case $N_\text{RIS} >1$ is given by
\begin{equation}
{\cal R}(\bm \Phi) = \log_2 \left( 1 + \SNR \left|h_\text{d} + \sum_{i = 1}^{N_\text{RIS}} [\bh_2]_i e^{\sfj \phi_i} [\bh_1]_i \right|_2^2 \right),
\label{equation:spectral_efficiency_siso}
\end{equation}
in which $[\bm \Phi]_{i,i} = e^{\sfj \phi_i}$. From \eqref{equation:spectral_efficiency_siso}, it is clear that the optimum $\phi_i$, $1 \leq i \leq N_\text{RIS}$, is the phase-shift that makes every complex element in the inner summation have the same phase as $h_\text{d}$. Let $\arg(z)$ denote the operation of extracting the phase of a complex number $z$. Then, the optimal elements of $\bm \Phi $ are provided as:
\begin{equation}
\phi_{i}^\star = \arg(\mathbf{h}_d) - (\arg(\mathbf{h}_{2,i}) + \arg(\mathbf{h}_{1,i}))
\label{equation:opt_RIS_phases_SISO}
\end{equation}
The $\phi_{i}^*$ are selected to coherently combine phases maximizing the squared magnitude of the equivalent channel $h_\text{eq}^\text{SISO} = h_\text{d} + \bh_2 \bm \Phi \bh_1$, and in the case of $N_\text{RIS} = 1$, the element $\phi_i$ is chosen to coherently combine the NLOS and LOS paths. The solution in \eqref{equation:opt_RIS_phases_SISO} is very attractive due to its simplicity, but it has the drawback of not being easily extensible to the MISO scenario. As we'll see next, the MISO cost function given below in \eqref{equation:spectral_efficiency_general} cannot be decoupled for the different $N_\text{RIS}$ phase elements $\phi_i$.
\subsection{Single-User MISO with single RIS element}
Considering a single RIS element in the SU-MISO setting yields a unique result. In this case, the beamformer $\bff \in \mathbb{C}^{N_\text{t} \times 1}$ in \eqref{equation:h_eq} needs to be optimized as well, whose solution is given by the maximum ratio transmission (MRT) beamformer \cite{Heathbook}. Similar to the SU-SISO scenario with both a single RIS element and multiple RIS elements in which the squared magnitude of the equivalent channel ${h_\text{eq}^\text{SISO}}$ was maximized, rate optimization in the MISO setting with a single RIS elements leads to optimization of the squared $\ell_2$-norm of the equivalent channel $\bh_\text{eq}^\text{MISO}$, given by
\begin{equation}
\begin{split}
	{\cal R}(\bff,\bm \Phi) &\propto  \Real\left\{ h_2 \bh_1 \bh_\text{d} e^{\sfj \phi} \right\} \\
	&= \left|h_2 \bh_1 \bh_\text{d} \right| \Real\left\{ e^{\sfj (\phi + \arg(h_2 \bh_1 \bh_\text{d})} \right\},
\end{split}
	\label{equation:MISO_single_RIS_pectral_efficiency_general}
\end{equation}
Here, $\phi$ is the RIS element phase to be configured, $h_2 \in \mathbb{C}^{\text{1x1}}$ is the channel scalar gain between the single RIS element and the UE, $\mathbf{h}_d \in \mathbb{C}^{1 \times N_\text{t}}$ is the line of sight (LOS) path between the BS and UE, $\mathbf{h}_1 \in \mathbb{C}^{1 \times N_\text{BS}}$ is the MISO channel between the single element RIS and the BS. The above optimization yields the optimal $\phi^\star = -\arg(h_2 \bh_1 \bh_\text{d})$. This scenario is similar to that of a single element relay-assisted wireless communication system \cite{RIS_myths,RIS_vs_Relays}, which is interesting for theoretical development, but less for practical purposes. 
\subsection{Single-User MISO with multiple RIS elements}
The scenario of more practical relevance is the SU-MISO scenario. Unlike the previously considered scenarios, this scenario does not allow for an optimal simple closed-form solution for the phases of the RIS elements. The optimization problem can be formally stated as follows:
\begin{equation}
\begin{split}
&\operatorname*{arg\,max}_\phi (\log_2\left( 1+(\SNR \mathbf{h}_\text{eq} \bff \bff^H \bh_\text{eq}^H \right)) \\
&\text{subject to } \left|\bee_i^T \bm \Phi \bee_i \right| = 1, \quad \forall i = 1:N_\text{RIS}
\end{split}
\end{equation}
Here, $\mathbf{h}_{eq}$ is defined in \eqref{equation:h_eq}. The challenge posed by this scenario is handling the non-convex objective function in conjunction with the non-convex unit-magnitude constraints on each of the $N_\text{RIS}$ RIS elements. This constraint enforces the phase shifting operation of the RIS. Unlike relays, RIS do not amplify or decode-then-forward a received signal, as such RIS elements require unit magnitude \cite{discrete_phase_SU_RIS, finite_element_iteration_paper, SDR_paper,mu_rate_optimization_tHz, mu_sum_rate_miso_opt_RIS, multiuser_RIS_problem,achievable_rate_opt_MIMO,cell-edge-user_ofdm_rate_max}, which may be relaxed theoretically to achieve an approximate solution. Similar to the optimization in the prior scenarios, prior work optimizes the $||\mathbf{h}_\text{eq}||^2$, leading to suboptimal solutions. In the following section, we propose rate optimization schemes that either maximize $||\mathbf{h}_\text{eq}||^2$ or directly through the spectral efficiency. 
\section{Proposed Approaches}
The proposed approaches can be categorized by the objective function used for optimization and the general class of methods used- power method and gradient-based approaches. These algorithms are summarized in Table \ref{tab:tradeoffs_algs}.
\begin{table*}[t!]
  \centering
  \caption{Trade-offs between prior work and proposed algorithms}
    \begin{tabular}{llllll}
		\toprule
		\textbf{Algorithm} & \textbf{Location} & \textbf{Rate} & \textbf{Complexity} & \textbf{Cost function} & \textbf{Iterative} \\
		Power method & See Alg \ref{alg:power method} & Medium & Medium & Equivalent channel power& Yes \\
		Gradient Ascent magnitude & See Alg \ref{alg:grad_method_magnitude_phases} & Medium & Medium & Equivalent channel power & Yes \\	
		Gradient Ascent magnitude RIS angles & See Alg \ref{alg:grad_method_magnitude_x} & Medium & Medium & Equivalent channel power & Yes \\	
		Gradient Ascent Spectral Efficiency & See Alg \ref{alg:grad_method_se} & High & Low & Spectral efficiency & Yes \\
		Gradient Ascent Spectral Efficiency power method initialization & See Alg \ref{alg:grad_method_se} & High & Medium & Spectral efficiency & Yes \\
		Fixed point Iteration & See \cite{finite_element_iteration_paper} & Low/Medium & High & Equivalent channel power& Yes \\
		Semi-Definite Relaxation & See \cite{SDR_paper} & Low & Low & Equivalent channel power & No \\
    \bottomrule
    \end{tabular}%
  \label{tab:tradeoffs_algs}%
\end{table*}%

\subsection{Optimization of magnitude of equivalent channel}
The optimization problem is formulated as follows
\begin{equation}
\begin{split}
\max &||\mathbf{h}_{\text{eq}}||_{2}^{2}\\
&\text{subject to } \left|\bee_i^T \bm \Phi \bee_i \right| = 1, \quad \forall i = 1:N_\text{RIS}\\
\label{equation:optimization_magnitude}
\end{split}
\end{equation}
The non-convex unit modal constraints in the optimization problem motivate the need for suboptimal tractable solutions to the rate optimization. The first approach to solve \eqref{equation:optimization_magnitude} consists of an application of the power method. The power method is an iterative method which converges to the eigenvector corresponding to the strongest eigenvalue of a matrix. We consider the power method as a low complexity algorithm consisting of the repeated application of a diagonalizable matrix to a vector iteratively. Let $\mathbf{H}_2 = \diag \left( \mathbf{h}_2\right)$, we consider the hermitian symmetric matrix $\mathbf{C} = \left(\mathbf{H}_{2} \mathbf{H}_{1} \right) {\left(\mathbf{H}_{2} \mathbf{H}_{1} \right )}^H$ in Algorithm 1 listed below. We let  $\mathbf{C}$ is contained in the expansion of the objective function $||\mathbf{h}_{\text{eq}}||_{2}^{2}$ where the diagonal $\mathbf{\Phi}$ is replaced with $\mathbf{H}_{2}$ and $\mathbf{h}_{2}$ . We also define a vector $\mathbf{b}$ as $\mathbf{H_2 H_1 h_d^H}$ and scalar $a$ as $\mathbf{h_{d} h_{d}^H}$.
\begin{equation}
\begin{split}
||\mathbf{h}_{\text{eq}}||_{2}^{2} = a +\mathbf{b}^H \mathbf{x} + \mathbf{x}^H \mathbf{b} + \mathbf{x}^{H}\mathbf{C} \mathbf{x}
\label{equation:h_eq_mag_opt}
\end{split} 
\end{equation}
In \eqref{equation:h_eq_mag_opt}, $\mathbf{x}$ corresponds to the vectorization of $\mathbf{\Phi}^c$. The application of $\mathbf{C}$ over $k$ iterations leads to $\mathbf{x}_{k}$ being a linear combination of the eigenvectors of $\mathbf{C}$ with corresponding eigenvalues raised to the $k$th power. As $k$ tends to infinity,  $\mathbf{x}$ tends to $\lambda_0^k  x_0 \mathbf{u_0}$; hence $\mathbf{x}_k$ points in the direction of the eigenvector $\mathbf{u}_0$ associated with the largest eigenvalue $\lambda_0$ of $\mathbf{C}$. The direction of $\mathbf{x}$ is of primary significance not the magnitude, as such we normalize $\mathbf{x}$ after each iteration to preserve unit length of $\mathbf{x}$ in order to avoid $\mathbf{x}$ from growing or shrinking without bound. Additionally we assume the maximum ratio transmisssion beamformer of form $\mathbf{f} = \mathbf{h}_{\text{eq}}^H/||\mathbf{h}_{\text{eq}}||_2$. 
\begin{algorithm}[!t]
\caption{Power Method for optimizing $||\mathbf{h}_{\text{eq}}||_{2}^{2}$}\label{alg:power method}
\begin{algorithmic}[1]
\State \textbf{Initialize $\mathbf{\Phi}$ randomly from a normal distribution} 
\State \textbf{Let $\mathbf{x}_0 = \text{vec} \left({\mathbf{\Phi}}\right)$} 
\State \textbf{Let} $\bC = \left(\bH_2 \bH_1\right) \left(\bH_2 \bH_1\right)^H$
\State \textbf{Set} $n = 1$
\While {$n < \text{maxIterations}$}
\vspace*{.15mm}
\qquad \State \textbf{Update $\mathbf{x}_{n}$ as $\mathbf{x}_{n} = \mathbf{C}\mathbf{x}_{n}$}
\vspace*{.15mm}
\qquad \State \textbf{Normalize $\mathbf{x}_{n}$ as $\mathbf{x}_{n} = \frac{\mathbf{x}_{n}}{||\mathbf{x}_{n+1}||_2}$}
\vspace*{.15mm}
\vspace*{.15mm}
\qquad \State \textbf{Evaluate objective function as $y_{n}$ using $\mathbf{x} = \mathbf{x}_{n}$}
\qquad \State \textbf{if $y_{n} > y_{n-1}$, then update $\mathbf{x}^\star = \mathbf{x}_{n}$} 
\qquad \State \textbf{Iteration update} as $n = n + 1$
\EndWhile
\State \textbf{Evaluate the ${\cal R}(\bff,\bm \Phi)$ with $\mathbf{x}^\star$ }  
\end{algorithmic}
\end{algorithm}

We observed that a random initialization of $\mathbf{x}$ was sufficient. We investigated initializing $\mathbf{x}$ with the solution to the linear system of equations $\mathbf{Cx} = \mathbf{b}$, and did not observe significant performance difference between that of $\mathbf{x}$ being randomly initialized. \\
\indent To optimize $||\mathbf{h}_\text{eq}||_2^2$, we also consider two gradient-based approaches in Algorithms \ref{alg:grad_method_magnitude_x} and\ref{alg:grad_method_magnitude_phases}. In Algorithm \ref{alg:grad_method_magnitude_phases}, we seek to optimize over the phases of $\mathbf{x}$ directly. In order to define the necessary gradient, we define $\mathbf{A}$ as $\diag{e^{-j\angle{\mathbf{x}}}}$, where $\angle{\mathbf{b}}$ is this initial set of RIS phases considered. The gradient of \eqref{equation:grad_mag_phases} with respect to the phases of $\mathbf{x}$ is
\begin{equation}
\nabla_{\angle{\mathbf{x}}} ||\mathbf{h}_{\text{eq}}||_{2}^{2}  = \left(\mathbf{b} + \mathbf{Cx}\right)^c \mathbf{A}
\label{equation:grad_mag_phases}
\end{equation}

In Algorithm \ref{alg:grad_method_magnitude_x} we perform gradient ascent on $\mathbf{x}$, rather than on  $\angle{\mathbf{x}}$ and define the corresponding gradient below:

\begin{equation}
\nabla_{\mathbf{x}} ||\mathbf{h}_{\text{eq}}||_{2}^{2}  = \left(\mathbf{b} + \mathbf{Cx}\right)^c
\label{equation:grad_mag_x}
\end{equation}
We enforce the unit magnitude constraint on each RIS elements in $\mathbf{x}$ by projecting the solution at each iteration onto the feasible set by considering only the phases of the updated $\mathbf{x}$ with unit magnitude.
\begin{equation}
\mathbf{x} = e^{j \angle{\mathbf{x}}}
\end{equation}
All iterative methods use a constant learning rate $\mu=.01$; a decaying learning rate can be considered in future work in efforts to enhance the gradient-based approach performances. Additionally, we consider a convergence threshold $\epsilon = 10^{-10}$ for both methods. \\

\begin{algorithm}[!t]
\caption{Projected Gradient Ascent over $\mathbf{x}$ for optimizing $||\mathbf{h}_{\text{eq}}||_{2}^{2}$}\label{alg:grad_method_magnitude_x}
\begin{algorithmic}[1]
\State \textbf{Let $\mathbf{\theta_{0}} = \angle{\left(\mathbf{b}\right)}$} 
\State \textbf{Let $\mathbf{x_{0}} = \mathbf{e}^{j \mathbf{\theta_{0}} }$} 
\State \textbf{Evaluate objective function $y_{0}$ in \eqref{equation:h_eq_mag_opt} using $\mathbf{x_{0}}$} 
\State \textbf{Set} $n = 1$
\While {$n < \text{maxIterations}$}
\vspace*{.15mm}
\State \textbf{Calculate  $\nabla_{\mathbf{x}_{n-1}} y_{n-1}$ according to \eqref{equation:grad_mag_x}}
\vspace*{.15mm}
\qquad \State \textbf{Update $\mathbf{x}_{n}$ as $\mathbf{x}_{n} = \mathbf{x}_{n-1} + \mu \nabla_{\mathbf{x}_{n-1}} y_{n-1} $}
\vspace*{.15mm}
\qquad \State \textbf{Let $\mathbf{x}_{n} = \mathbf{e}^{j \theta_{n}}$}
\vspace*{.15mm}
\qquad \State \textbf{Evaluate objective function as $y_{n}$ using $\mathbf{x} = \mathbf{x}_{n}$}
\qquad \State \textbf{if $||y_{n} - y_{n-1}||_2 \leq \epsilon$, then convergence met; break} 
\qquad \State \textbf{Iteration update} as $n = n + 1$
\EndWhile
\State \textbf{Evaluate ${\cal R}(\bff,\bm \Phi)$ according to \eqref{equation:spectral_efficiency_general} with $\mathbf{x}^\star$ }  
\end{algorithmic}
\end{algorithm}

\begin{algorithm}[!t]
\caption{Projected Gradient Ascent over $\angle{\mathbf{x}}$ for optimizing $||\mathbf{h}_{\text{eq}}||_{2}^{2}$}\label{alg:grad_method_magnitude_phases}
\begin{algorithmic}[1]
\State \textbf{Let $\mathbf{\theta_{0}} = \angle{\left(\mathbf{b}\right)}$} 
\State \textbf{Let $\mathbf{x_{0}} = \mathbf{e}^{j \mathbf{\theta_{0}} }$} 
\State \textbf{Evaluate objective function $y_{0}$ in \eqref{equation:h_eq_mag_opt} using $\mathbf{x_{0}}$}  
\State \textbf{Set} $n = 1$
\While {$n < \text{maxIterations}$}
\vspace*{.15mm}
\State \textbf{Define $\mathbf{A}$ as $\diag{\left(-j \mathbf{\theta}_{n-1} \right)}$}
\vspace*{.15mm}
\State \textbf{Calculate  $\nabla_{\angle{\mathbf{x}_{n-1}}} y_{n-1}$ according to \eqref{equation:grad_mag_phases}}
\vspace*{.15mm}
\qquad \State \textbf{Update $\mathbf{\theta}_{n}$ as $\mathbf{\theta}_{n} = \mathbb{R} \left(\mathbf{\theta}_{n-1} + \mu \nabla_{\angle{\mathbf{x}_{n-1}}} y_{n-1} \right) $}
\vspace*{.15mm}
\qquad \State \textbf{Let $\mathbf{x}_{n} = \mathbf{e}^{j \theta_{n}}$}
\vspace*{.15mm}
\qquad \State \textbf{Evaluate objective function $y_{n}$ in \eqref{equation:h_eq_mag_opt} using $\mathbf{x} = \mathbf{x}_{n}$}
\qquad \State \textbf{if $||y_{n} - y_{n-1}||_2 \leq \epsilon$, then convergence met; break} 
\qquad \State \textbf{Iteration update} as $n = n + 1$
\EndWhile
\State \textbf{Evaluate the ${\cal R}(\bff,\bm \Phi)$ with $\mathbf{x^*}$ } 
\end{algorithmic}
\end{algorithm} 

\subsection{Optimization directly over rate}

Majority of prior work tries to optimize \eqref{equation:h_eq_mag_opt}. Important contributions of this paper are twofold: i) to optimize the problem in \eqref{equation:opt_problem_spectral_efficiency} directly, and ii) to determine if there is a performance difference between optimizing over ${\cal R}(\bff,\bm \Phi)$ directly versus optimizing over $||\mathbf{h}_\text{eq}||_2^2$ in \eqref{equation:h_eq_mag_opt}. This is an important analysis because in previous application cases an optimal closed-form solution for the elements of $\mathbf{\Phi}$ could be obtained; however in the SU-MISO with multiple RIS elements case-- in which an optimal solution is not available-- such an analysis is pertinent particular for future extensions into the MIMO case and beyond.

The proposed algorithm is a projected gradient ascent over ${\cal R}(\bff,\bm \Phi)$. We derive $\nabla_{\mathbf{\Phi}} {\cal R}(\bff,\bm \Phi)$, the gradient of ${\cal R}(\bff,\bm \Phi)$ with respect to $\mathbf{\Phi}$ below: 

\begin{equation}
\begin{split}
{\cal R}(\bff,\bm \Phi) = \log_{2}{\bigg{(}} 1 + \SNR &{\big{(}}\bh_\text{d} \bff \, \bff^H \bH_1^H \bm\Phi^H \} \bh_2^H \\ 
& + \bh_\text{d} \bff \, \bff^H \bh_\text{d}^H \\
& + \bh_2 \bm\Phi \bH_1 \bff \, \bff^H \bH_1^H \bm\Phi^H \bh_2^H \\ 
& + \bh_2 \bm\Phi \bH_1 \bff \, \bff^H \bh_\text{d}^H {\big{)}} {\bigg{)}}
\end{split}
\label{equation:expanded_se}
\end{equation}

\noindent Taking the derivative of each term within $\log_2(\cdot)$: 

\begin{align*}
\frac{\partial \bh_\text{d} \bff \, \bff^H \bH_1^H \bm\Phi^H \bh_2^H}{\partial \mathbf{\Phi}} = 0 \quad  \frac{\partial \bh_\text{d} \bff \, \bff^H \bh_\text{d}^H}{\partial \mathbf{\Phi}} = 0
\end{align*}

\begin{align*}
\frac{\partial \bh_2 \bm\Phi \bH_1 \bff \, \bff^H \bH_1^H \bm \Phi^H \bh_2^H}{\partial \mathbf{\Phi}} = \frac{\partial \mathbf{g \Phi d}}{\partial \mathbf{\Phi}} \\
\end{align*}

\noindent where $\mathbf{g}  \in \mathbb{C}^{1 \times N_\text{RIS}}$ and $\mathbf{d} \in \mathbb{C}^{N_\text{RIS} \times 1}$

\begin{align*}
 \frac{\partial \mathbf{g \Phi d}}{\partial \mathbf{\Phi}} &= \frac{\partial \sum_{j = 1}^{N_\text{RIS}} \left[ \sum_{i = 1}^{N_\text{RIS}} \mathbf{g}_{i} \mathbf{I_{:,i}} \right]_{j} \mathbf{d}_{j}}{\partial \mathbf{\Phi}_{i,j}} \\
 &= \left(\mathbf{g} \otimes \mathbf{d}^T \right)^T
\end{align*}

\noindent Similarly we define the partial derivative of the last term within the $\log_2(\cdot)$ in \eqref{equation:expanded_se}:

\begin{align*}
\frac{\partial \bh_2 \bm\Phi \bH_1 \bff \, \bff^H \bh_\text{d}^H}{\partial \mathbf{\Phi}} = \frac{\partial \mathbf{k \Phi s}}{\partial \mathbf{\Phi}}
\end{align*}

\noindent where $\mathbf{k}  \in \mathbb{C}^{1 \times N_\text{RIS}}$ and $\mathbf{s} \in \mathbb{C}^{N_\text{RIS} \times 1}$

\begin{align*}
 \frac{\partial \mathbf{g \Phi d}}{\partial \mathbf{\Phi}} = \left(\mathbf{k} \otimes \mathbf{s}^T \right)^T
\end{align*}

\noindent Taking the derivative of $\log_2(\cdot)$ and applying the chain rule with the above derivatives gives the following full gradient:

\begin{equation}
\begin{split}
\nabla_{\mathbf{\Phi}} & {\cal R}(\bff,\bm \Phi) = \\
&\frac{\SNR \left[ (\bH_1 \bff \, \bff^H \bH_1^H \bm \Phi^H \bh_2^H \otimes \bh_2 \right)^T + (\bH_1 \bff \, \bff^H \bh_3^H \otimes \bh_2)^T ]}{1 + \SNR \bh_\text{eq} \bff \, \bff^H \bh_\text{eq}^H} \\
\end{split}
\label{equation:grad_se}
\end{equation}

\begin{algorithm}[!t]
\caption{Projected Gradient Ascent optimizing ${\cal R}(\bff,\bm \Phi)$}\label{alg:grad_method_se}
\begin{algorithmic}[1]
\State \textbf{Initialize $\mathbf{\Phi}$ with random phases taken from ${\cal N} (0,1)$
\State \textbf{Initialize the precoder $\mathbf{f}$ as $\mathbb{C}^{N_{\text{t} \times 1}}$}
\State \textbf{Normalize $\mathbf{f}$ such that $||\mathbf{f}||_2 =1$}}
\State \textbf{Evaluate ${\cal R}_0(\bff,\bm \Phi)$ according to \eqref{equation:spectral_efficiency_general} using $\bff$, $\bm\Phi$} %
\State \textbf{Set} $n = 1$
\While {$n < \text{maxIterations}$}
\vspace*{.15mm}
\State \textbf{Calculate  $\nabla_{\mathbf{\Phi_n}} {\cal R}(\bff,\bm \Phi)$ according to \eqref{equation:grad_se}}
\vspace*{.15mm}
\qquad \State \textbf{Normalize magnitude of diagonal elements of $\mathbf{\Phi}_n$}
\vspace*{.15mm}
\qquad \State \textbf{Compute $\mathbf{h}_{\text{eq},n} = \mathbf{h}_\text{d} + \mathbf{h}_2 \mathbf{\Phi}_n \mathbf{H}_{1}$}
\vspace*{.15mm}
\qquad \State \textbf{Update $\bff_n$ as $\bh_{\text{d},n}^H/\|\bh_{\text{d},n}\|_2$}  
\vspace*{.15mm}
\qquad \State \textbf{Evaluate ${\cal R}_{n}(\bff,\bm \Phi)$ by \eqref{equation:expanded_se} using $\bff_n$, $\bm\Phi_n$}
\qquad \State \textbf{if} $|{\cal R}_{n}(\bff,\bm \Phi) - {\cal R}_{n-1}(\bff,\bm \Phi)| \leq \epsilon$ \textbf{then}
\qquad \qquad \hspace*{.25in} \textbf{Convergence and break} 
\qquad \State \textbf{Iteration update as $n = n + 1$}
\qquad \State \textbf{Update learning rate $\mu$ as $\mu  = \frac{\mu}{2}$}
\EndWhile
\State \textbf{Evaluate the optimized ${\cal R}(\bff^\star,\bm \Phi^\star)$ according to \eqref{equation:spectral_efficiency_general}} 
\end{algorithmic}
\end{algorithm} 

Similar to prior work \cite{finite_element_iteration_paper, SDR_paper} and the other proposed algorithms, we employ a MRT beamformer at each iteration to configure $\bff$. Unlike Algorithms \ref{alg:grad_method_magnitude_phases} and \ref{alg:grad_method_magnitude_x}, a decaying $\mu$ is utilized in order to facilitate convergence. We initialize $\mu$ to $0.5$ We also investigated initializing Algorithm \ref{alg:grad_method_se} with the power method solution in comparison to random initialization and observed a performance difference which is shown in the following section. Additionally, we observed reinitializing the learning rate after the learning rate had decayed sufficiently-- in order to escape a potential local optimal-- did not lead to substantial performance difference between that of before reinitialization. The proposed projected gradient ascent method is detailed in Algorithm \ref{alg:grad_method_se}.


\section{Numerical Results}
In this section, we present numerical results on both spectral efficiency and computational complexity for both the different proposed algorithms and those of prior work. We considered simulations over $N_\text{MC} = 100$ Monte Carlo realizations using the signal model in \eqref{equation:rx_signal_general}. We consider the case in which the transmitting BS is equipped with $\Nt = 32$ antennas, we analyze the scenario in which both the BS and RIS have the same number of corresponding active and passive elements respectively.

\subsection{Spectral Efficiency Analysis}

In our analysis we make a comparison with two algorithms from prior work which aim to optimize the spectral efficiency in a MISO RIS-assisted communication system \cite{finite_element_iteration_paper, SDR_paper}. These particular algorithms were selected as appeared to be most relevant to the single user scenario considered. Furthermore, they are regarded as state-of-the-art solutions according to in \cite{overview_RIS_journal}. 

\begin{figure}
    \centering
    \includegraphics[width=0.9\columnwidth]{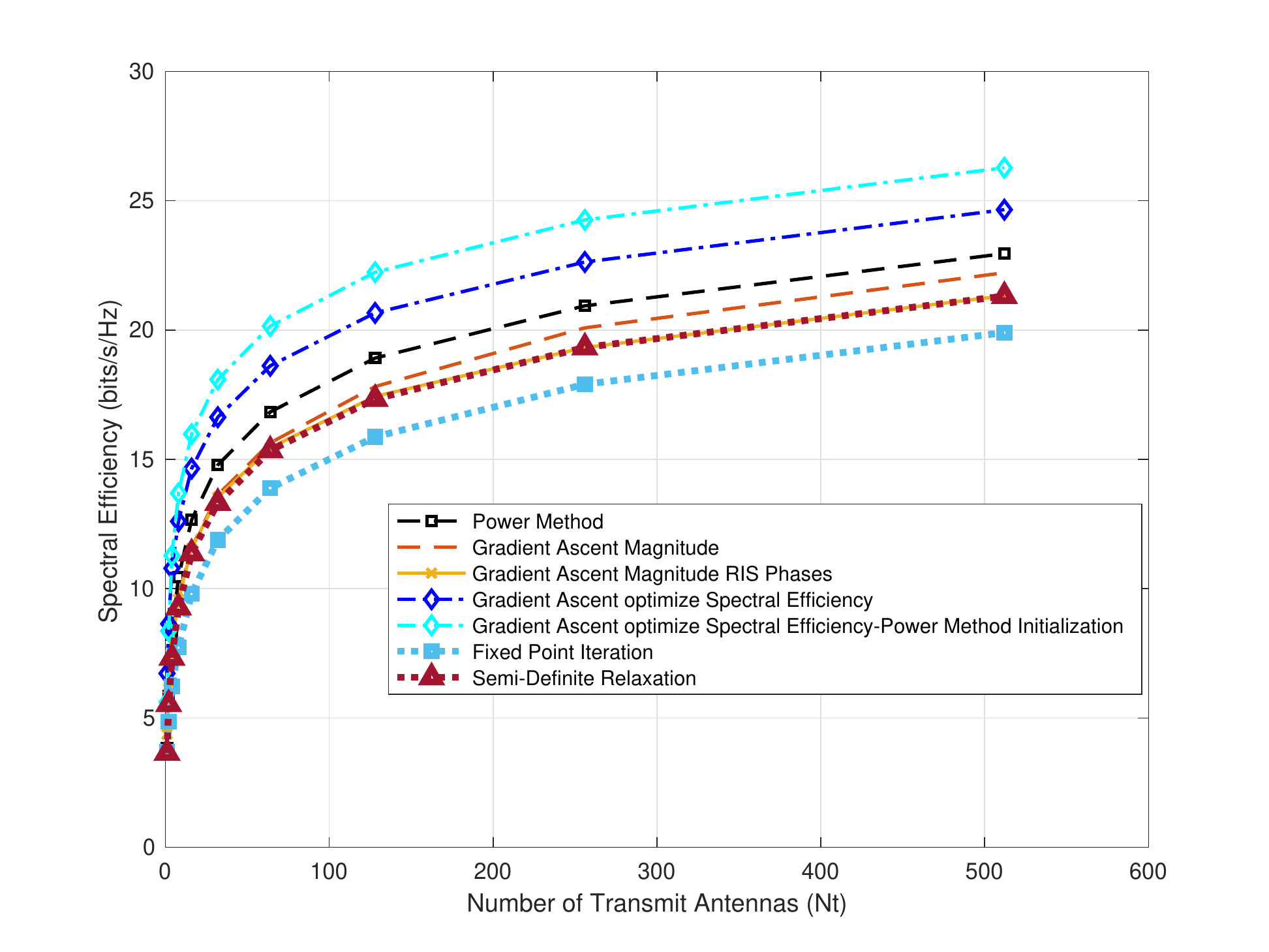}
    \caption{Spectral efficiency vs. $N_{\text{t}}$ for $\SNR = 10$ dB.}
    \label{fig:se_vs_nt}
\end{figure}

\begin{figure}
    \centering
    \includegraphics[width=0.9\columnwidth]{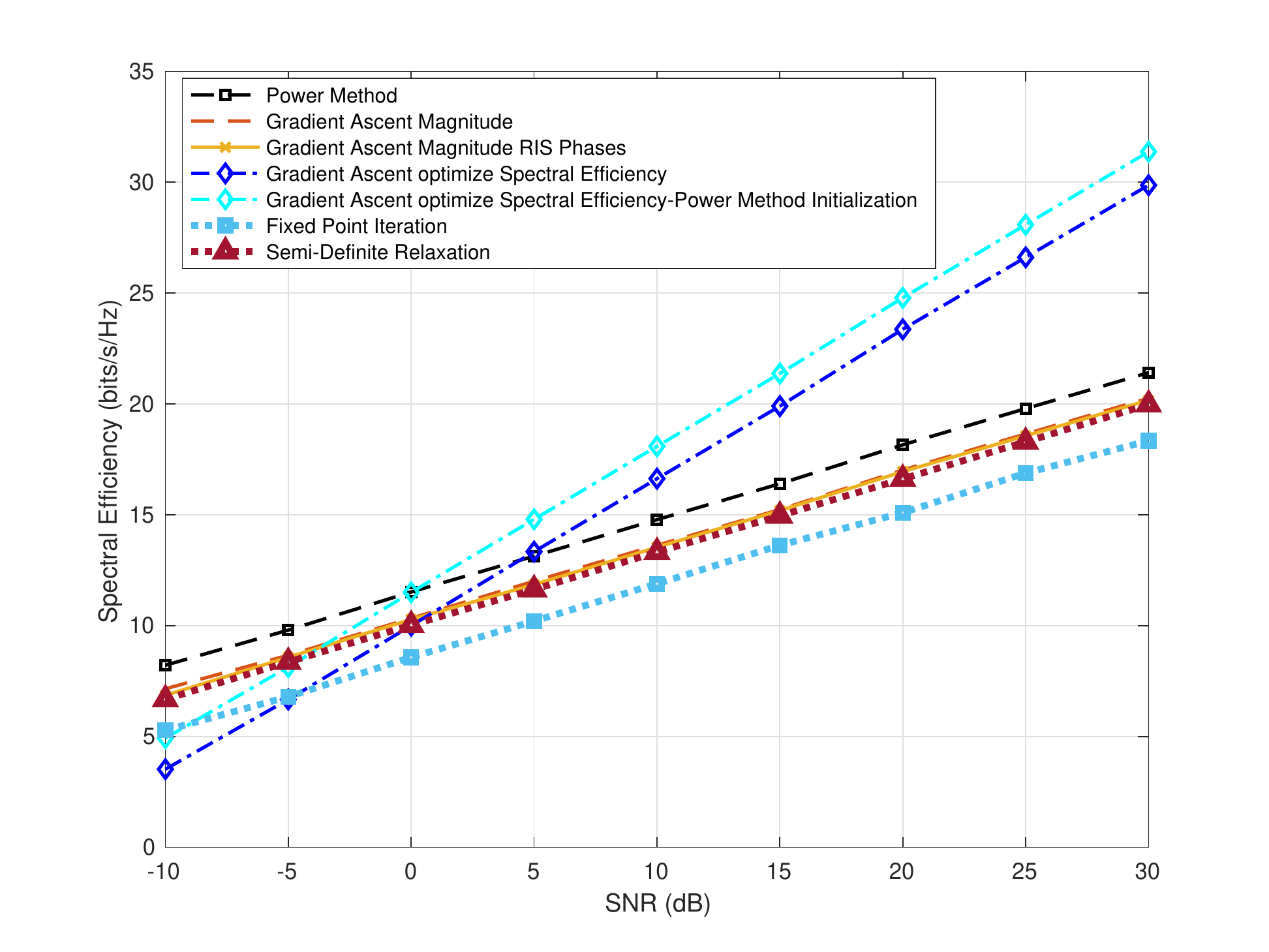}
    \caption{Spectral efficiency vs. $\SNR$ for $N_{\text{t}} = N_{\text{RIS}} = 32$.}
    \label{fig:se_vs_snr}
\end{figure}

From Fig. \ref{fig:se_vs_nt} and \ref{fig:se_vs_snr} we can observe how the spectral efficiency performance of proposed algorithms and comparison algorithms evolve as $N_\text{t}$ and SNR increase. We observe that Algorithm \ref{alg:grad_method_magnitude_phases} is the only Algorithm which does not outperform prior work. We observe a similar behavior for different values of $\SNR$ or $N_\text{t}$. The Algorithms \ref{alg:grad_method_magnitude_x} and \ref{alg:grad_method_magnitude_phases} have similar spectral efficiency performance when $N_\text{t}$ and $N_\text{RIS}$ are small, but as $N_\text{t}$ and $N_\text{RIS}$ increase, the performance of Algorithm \ref{alg:grad_method_magnitude_phases} decays to the performance level of the SDR algorithm from \cite{SDR_paper}; while the Algorithm \ref{alg:grad_method_magnitude_x} maintains higher performance than both comparison algorithms. As such, we determine it is a better design choice to conduct gradient ascent over $\bx$ rather than $\angle \bx$ when optimizing \eqref{equation:h_eq_mag_opt}. Furthermore, in Figs. \ref{fig:se_vs_nt} and \ref{fig:se_vs_snr} we observe that optimization directly over the spectral efficiency cost function yields the highest performance gains over the other proposed algorithms. Using the solution from Algorithm \ref{alg:power method} as the initialization to algorithm \ref{equation:h_eq} unlocks further spectral efficiency gains over the base Algorithm \ref{alg:grad_method_se}.


\subsection{Complexity Analysis}

\begin{figure}
    \centering
    \includegraphics[width=0.9\columnwidth]{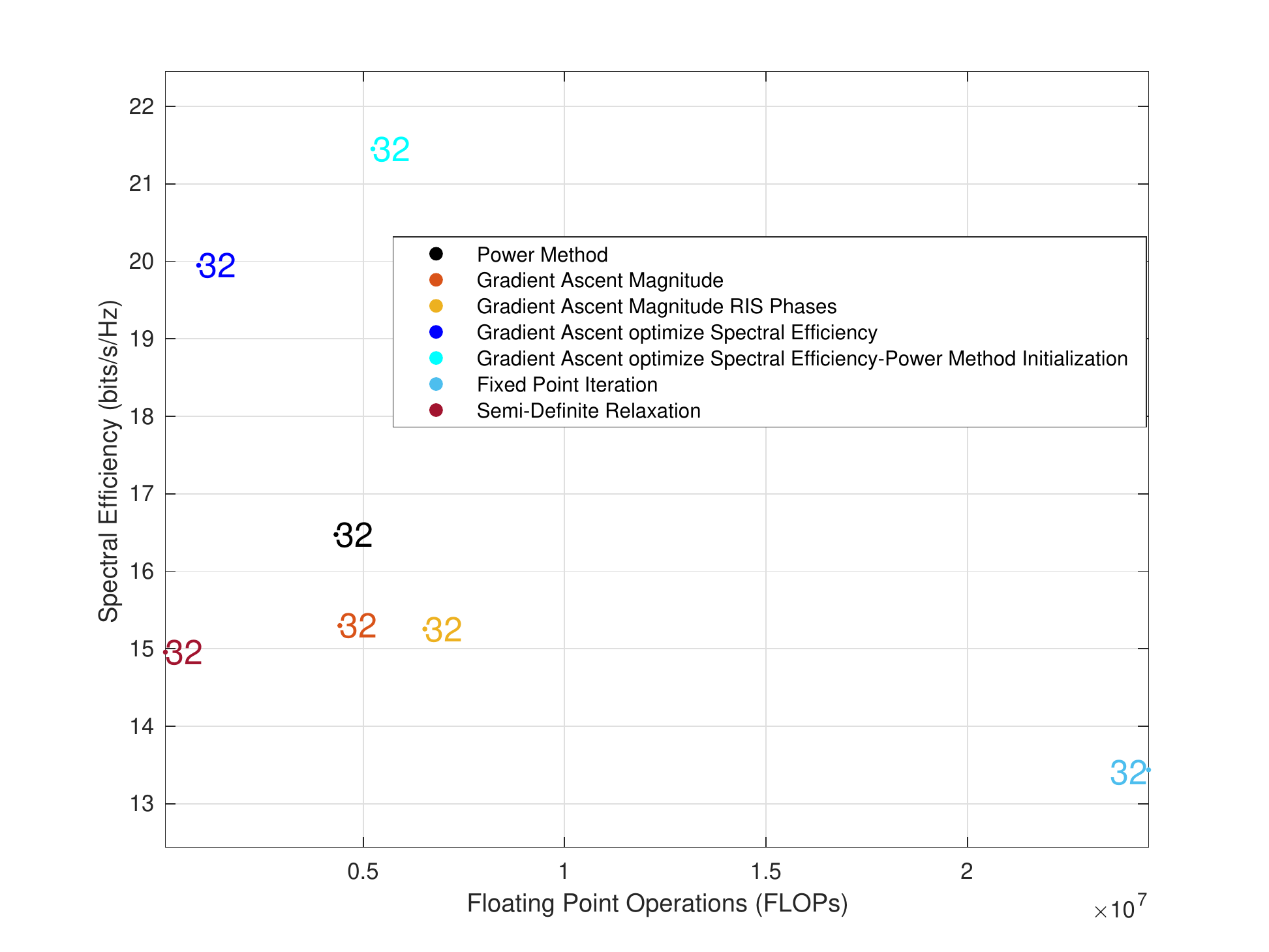}
    \caption{Spectral efficiency vs. floating point operations (FLOPs) for $N_{\text{t}} = N_{\text{RIS}} = 32$. The marker used indicate the size of the BS antenna array, as well as the RIS panel size.}
    \label{fig:se_vs_flops}
\end{figure}

Another critical criterion for evaluating algorithms includes the computational complexity \cite{achievable_rate_opt_MIMO,overhead_aware_RIS_configuration}. We quantify the runtime complexity as the number of Floating Point Operations (FLOPs), and obtain the average spectral efficiency and average number of FLOPs in a Pareto scatter plot in Fig. \ref{fig:se_vs_flops}; furthermore Fig. \ref{fig:se_vs_flops} allows understanding the Pareto frontier of the algorithms investigated within this paper.  We aim to generate an algorithm which is characterized by high spectral efficiency performance and low complexity. We observe that, for the given experimental scenario, Algorithm \ref{alg:grad_method_se} provides relatively high achievable spectral efficiency and relatively low computational complexity. Algorithm \ref{alg:grad_method_se} using the power method initialization yields slightly higher spectral efficiency at the cost of multiplying the complexity of the algorithm by a factor of $4$ approximately. Algorithms \ref{alg:power method}, \ref{alg:grad_method_magnitude_phases}, and \ref{alg:grad_method_magnitude_x} are a cluster of algorithms classified by similar middle range performance in terms of spectral efficiency and complexity, however; amongst these middle range algorithms, Algorithm \ref{alg:power method} provides the highest spectral efficiency and lowest complexity. 

\section{Conclusion}
In this paper, we proposed multiple strategies for spectral efficiency optimization of a RIS-assisted MISO communication system. We determined performance gains dependent upon which cost function is exploited by the various algorithms and showed how the proposed algorithms perform favorably against the current state-of-the-art algorithms, both in terms of spectral efficiency and complexity. Furthermore, we drew insights regarding how spectral efficiency of the RIS-assisted MISO system is affected by the dimensionality of the optimization problem and $\SNR$ used. Overall, we observe that the proposed projected gradient ascent over spectral efficiency offers a promising trade-off between different performance metrics for the sub-$6$ GHz system considered in this paper.

\bibliographystyle{IEEEtran}

\end{document}